\documentclass[prb,twocolumn,superscriptaddress,showpacs]{revtex4}

\usepackage{amsmath}
\usepackage{amssymb}
\usepackage{amsfonts}

\usepackage{graphicx}
\usepackage{times}
\usepackage{pstricks}
\usepackage{tikz}
\usepackage{subfig}
\usepackage{comment}
\usetikzlibrary{patterns}

\newcommand{\be}{\begin{equation} }
\newcommand{\ee}{\end{equation} }
\newcommand{\ba}{\begin{eqnarray} }
\newcommand{\ea}{\end{eqnarray} }

\begin{document}


\title{Microwave Absorption by a Mesoscopic Quantum Hall Droplet}

\author{Jennifer Cano}
\affiliation{Department of Physics, University of California, Santa Barbara, CA 93106}
\author{Andrew C. Doherty}
\affiliation{ARC Centre of Excellence for Engineered Quantum Systems, School of Physics,
The University of Sydney, Sydney, NSW 2006, Australia}
\author{Chetan Nayak}
\affiliation{Station Q, Microsoft Research, Santa Barbara, CA 93106-6105}
\affiliation{Department of Physics, University of California, Santa Barbara, CA 93106}
\author{David J. Reilly}
\affiliation{ARC Centre of Excellence for Engineered Quantum Systems, School of Physics,
The University of Sydney, Sydney, NSW 2006, Australia}

\date{\today}

\begin{abstract}
We consider the absorption of microwaves by a quantum Hall droplet.
We show that the number and
velocities of charged edge modes can be directly measured from
a droplet of known shape. In contrast to standard transport measurements, different edge equilibration regimes can be accessed in the same device. If there is a quantum point contact in
the droplet, then quasiparticle properties, including braiding statistics,
can be observed. Their effects are manifested as modulations of
the microwave absorption spectrum that are, notably, first-order in the
tunneling amplitude at the point contact.
\end{abstract}

\maketitle


\section{Introduction}

A classic problem in mathematical physics asks ``can you hear the shape of a drum?''\cite{Kac,GordonWebbWolpert}
In this paper, we address the natural generalization: ``can you hear an anyon
in a drum?'' For the sake of concreteness, we consider a `drum' that is
a mesoscopic quantum Hall device of circumference $L \approx 10 -100\mu\text{m}$.
The excitations of the edge of a quantum Hall droplet, which are gapless
in the limit of a large droplet, have a minimum energy $2\pi \hbar v/L$,
where $v$ is the velocity of edge modes and $L$ is the circumference of the droplet.
Therefore, for $v=10^4 - 10^5\, \text{m}/\text{s}$, the edge modes of such
a drum can be `heard' in the frequency range $\approx 1-100\, \text{GHz}$ or,
in other words, with microwaves. Such modes have already been observed using spectroscopy\cite{Allen83,Talyanskii89a,Talyanskii89b,Wassermeier90,Andrei88,Grodnensky91,Talyanskii94,Kukushkin05,Smetnev11}  in samples on the millimeter scale and analyzed using semiclassical models,\cite{Volkov,Fetter} and have also been observed through time resolved measurements.\cite{KamataVelocity,Kumada11,AshooriVelocity,Zhitenev94,Ernst97} Here, we focus exclusively on the absorption spectrum of micron-scale samples tuned to quantum Hall plateaus.

As we show in this paper, microwave absorption gives a window
into edge excitations that is different from and complementary to
transport\cite{Wen90,Ernst97}. Moreover, it provides a probe that can enable one to
observe a key feature of the theory of fractional quantum Hall states -- the
exotic braiding statistics of its excitations -- that has, thus far,
remained somewhat elusive experimentally.
The fractional charge and statistics of quasiparticles
are the lynchpins of the theory of the fractional quantum
Hall effect. According to Laughlin's gauge argument,
fractional quantized Hall conductance can only occur
if there are quasiparticles with fractional charge \cite{Laughlin81}.
There is strong experimental support for fractional charge
$e^*=e/3$\cite{ObsFracChargeGoldmanSu,ObsFracChargeHeiblum,ObsFracChargeGlattli,ObsFracChargeYacoby,McClureInterferometry,CaminoInterferometry} at $\nu=1/3,7/3$
and for $e^*=e/4$\cite{QuarterChargeHeiblum,Willett09,QuarterChargeYacoby}
at $\nu = 5/2$.
But fractionally-charged quasiparticles
must have fractional braiding statistics \cite{Kivelson},
and both microscopic wavefunctions \cite{Halperin84,Arovas84}
and long-wavelength effective field theories \cite{Zhang89,Read89a} predict that
quasiparticles in the fractional quantum Hall effect are anyons.
However, the braiding properties
of quasiparticles are not directly manifested in bulk transport experiments
or even in transport through a quantum point contact.
A two point contact interferometer device is, until now, the only proposed
way to directly observe them. Although there is a measurement \cite{Willett09}
that is consistent with non-Abelian anyon quasiparticles at $\nu=5/2$, it is not definitive since it has not been reproduced and
other interpretations are conceivable. The setup described in
this paper would enable a truly distinct and independent measurement
of quasiparticle braiding properties. Moreover, it can enable the measurement of
some aspects of the physics of quantum Hall edge excitations, such
as the number of edge modes and their velocities, that are difficult to directly
observe in transport experiments.

Our proposed setup consists of a quantum Hall droplet or circular disk coupled to a broadband microwave co-planar waveguide and used as a microwave spectrometer, as shown in Fig.~\ref{fig:setup}.
The electric field between the central track and ground of the waveguide couples to the charged edge modes of the droplet and allows a non-invasive means of probing the system, without contacting the electron gas. The absorption spectrum, determined by measuring the amount of transmitted microwave power through the waveguide, will be one or more series of peaks corresponding to the allowed edge modes of the droplet.

For a circular droplet, there will be one peak for each charged edge mode
and the positions of the peaks in frequency provide a direct measurement of the velocity of the mode.
This is particularly interesting for certain fractions that are predicted to have counter propagating edge modes on a completely clean edge
but one charged and one oppositely propagating neutral mode on a disordered edge large enough for modes to equilibrate \cite{KaneFisherGeneral,KaneFisher23}. The most notable example is at $\nu = 2/3$, where such neutral modes were recently observed.\cite{ObserveNeutral,LocalThermometry} In our setup, we would expect to see
one charged mode in a device that is larger than the equilibration length,
and multiple modes in smaller devices. The latter possibility has not yet
been observed.
Furthermore, a surprising result in Ref.~\onlinecite{LocalThermometry} is the observation of an upstream mode at $\nu = 1$ and, simultaneously, a local Hall resistivity of $\frac{3}{2} \frac{h}{e^2}$. This observation indicates that a local measurement between points 20 $\mu$m apart is at a distance less than the equilibration length and in this region, the edge supports both $\nu = 1$ and $\nu = 2/3$ edges. Hence, in a droplet smaller than this length, one would detect multiple charged modes. At $\nu=5/2$, one of the candidate
states, the anti-Pfaffian state \cite{Lee07,Levin07} similarly has two phases
of edge excitations, one with a single charged mode and one with two, one
upstream and one downstream.

A quantum point contact (QPC) can be produced by fabricating standard surface gates that overlap the droplet to create an interferometer: the heights of the peaks oscillate as a function of magnetic field and the oscillations experience a phase slip when the number of quasiparticles changes. However, unlike in a
two-point contact interferometer, the oscillations are {\em first-order in the tunneling amplitude}.

In what follows, we first compute the absorption spectrum of a quantum Hall droplet in an integer or Laughlin state with no QPC. Next, we consider filling fractions with more complicated edges and show how the absorption spectrum reveals the number of current carrying modes. Then, we add the QPC and show how the spectrum acts as an interferometry measurement. We then repeat the calculation for filling fraction $\nu = 5/2$ and predict the non-Abelian interference pattern.

\section{General Considerations}
\label{sec:general}

\subsection{Experimental Setup}

		

\begin{figure}
\includegraphics[width=3.25in]{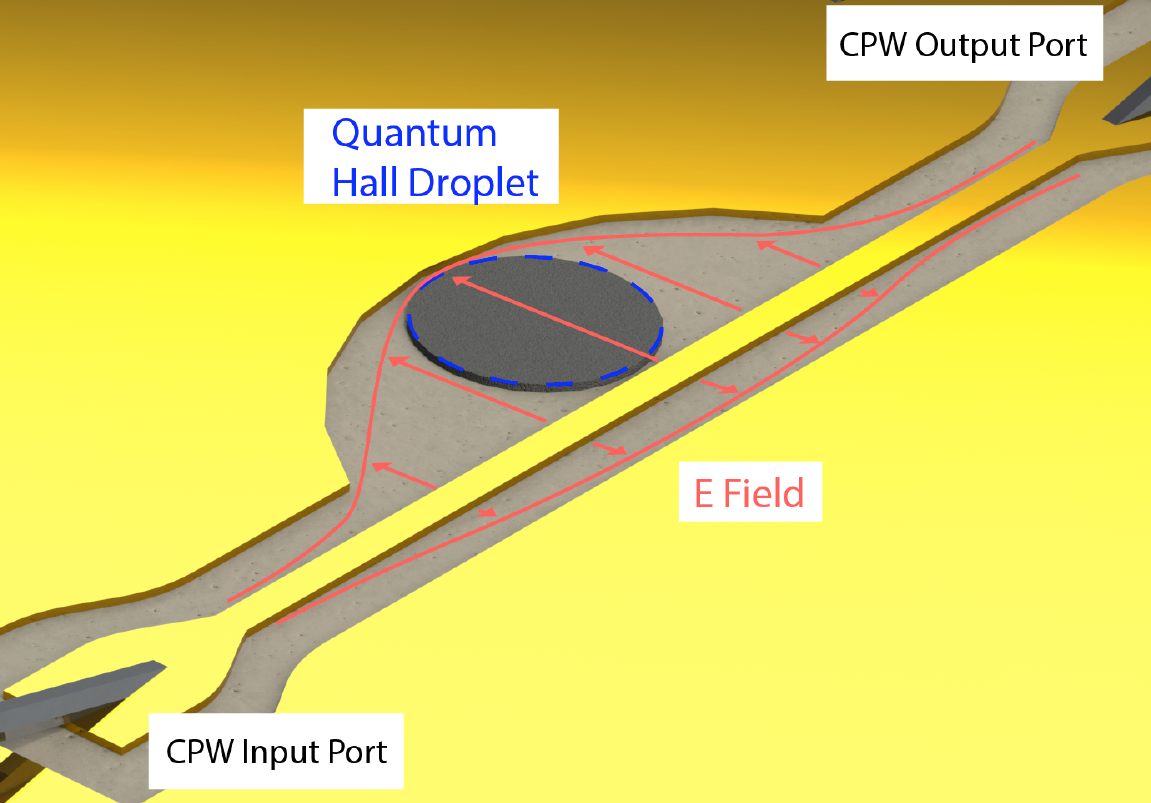}
	\caption{Quantum Hall droplet in a microwave waveguide. Note that the droplet and microwave wavelength are not drawn to scale. The latter is much larger, and the electric field is approximately constant on the scale of the droplet.}
\label{fig:setup}
\end{figure}

We consider a single mesoscopic disk or droplet of electron gas in the quantum Hall regime and capacitively coupled to a broadband, coplanar waveguide or narrow band resonator (see Fig.~\ref{fig:setup}). The droplet is not in ohmic contact with any leads, allowing continuous edge states around its circumference, and is probed purely through its microwave absorption spectrum. When the frequency of incident microwaves is tuned so that they are resonant with one of the excited states of the quantum Hall device, there will be absorption and a corresponding change in the amplitude and phase of the microwave power transmitted through the coplanar waveguide. Small changes in microwave  power and phase are readily measured with cryogenic amplification and standard homodyne detection techniques. We envision devices that also include an adjacent Hall-bar, fabricated on-chip but sufficiently far away to be decoupled from the droplet. This Hall-bar then enables simultaneous transport measurement for comparison with microwave data.

The microwave response for clean systems will comprise components from low-energy excitations in the bulk as well as edge magnetoplasmons. Between Landau levels (where the transverse Hall conductance shows plateaus in transport), the edge magnetoplasmons will dominate the microwave response over the bulk-excitations; assuming that the dipole matrix element for bulk-excitations is
$\sim e \ell_0$, where $l_0$ is the magnetic length, such states will give a weak background contribution to the absorption $R(\omega)$
that is $\sim \omega^2 (e \ell_0)^2$.
When the number of occupied Landau levels is changing however, bulk excitations can lead to significant absorption $R(\omega)\sim\Gamma/(\omega^2+\Gamma^2)$ characteristic
of a metallic state (where the DC conductivity is $\sim 1/\Gamma$). In this regime the contribution from edge magnetoplasmons will merge with the spectrum of bulk excitations. We note however, that these contributions can be easily separated by measuring the microwave response as a function of frequency and magnetic field. 

Of course, even on a plateau, if the microwave frequency is higher than
the mobility gap, then there will be absorption characteristic of a metal
in the bulk. Therefore, in order to be resonant with an edge excitation
of energy $\omega=2\pi v/L$ (where $L$ is the circumference of the droplet)
but still below the bulk mobility gap (or, operationally, the gap deduced from
transport, $\Delta_{\rm tr}$), we need $2\pi v/L < \Delta_{\rm tr}$. Thus,
relatively large devices and smaller velocities are advantageous. Large devices are also expected to couple more strongly to the electric field from the waveguide and hence show a larger response. On the other hand, in order to probe different equilibration regimes and to observe quantum
interference effects, it is advantageous to have smaller devices. 
Thus, there is an intermediate regime $L\approx 10 \mu\text{m}$ and
velocity $v\approx 10^4\,\text{m/s}$ in which we expect to be able to isolate
the physics of edge excitations if the system lies on a quantum Hall plateau.
Note that some experimental observations\cite{McClureVelocity,AshooriVelocity,KamataVelocity} are consistent with a larger velocity $\sim 10^5 \, \text{m/s}$; it may be necessary to tune gate potentials and the magnetic field in order to have a smaller velocity in which the aforementioned intermediate
regime of frequencies exists \cite{Smetnev11, Kumada11}.

\subsection{Kohn's Theorem}
\label{sec:Kohn}

As mentioned in the introduction, for a very clean quantum Hall device at a filling fraction with multiple edge modes, we expect to see one peak in the absorption spectrum for every charged edge mode with a distinct velocity. This is of particular interest for fractions predicted to have a disorder driven (equilibrated) fixed point, because the clean and disordered systems would have different numbers of charged modes, and hence different signatures in the absorption spectrum.

On the other hand, since the electric field is nearly constant on the scale
of the the quantum Hall device, it couples to the dipole moment of the
system as follows:
\begin{eqnarray}
H &=& {\sum_i}\frac{{\bf p}_i^2}{2m} + \sum_{i>j} V({{\bf r}_i}-{{\bf r}_j})
+ {\sum_i} U({{\bf r}_i}) + e{\sum_i}{{\bf r}_i}\cdot{\bf E}\cr
&=& \frac{{\bf P}_{\text{c.m.}}^2}{2Nm} + Ne{\bf R}_{\text{c.m.}}\cdot{\bf E}\cr
& & \,\, + \,\,{\tilde U}({\bf R}_{\text{c.m.}},{\bf r}_{\text{rel},1},\ldots,{\bf r}_{\text{rel},N-1})
\,+ \, H_{\text{rel}}
\end{eqnarray}
where ${\bf R}_{\text{c.m.}}$ and ${\bf P}_{\text{c.m.}}$ are the
center-of-mass coordinate and momentum;
${\bf r}_{\text{rel},1},\ldots,{\bf r}_{\text{rel},N-1}$ are the relative coordinates;
${\tilde U}({\bf R}_{\text{c.m.}},{\bf r}_{\text{rel},1},\ldots,{\bf r}_{\text{rel},N-1})\equiv
{\sum_i} U({{\bf r}_i})$; and $H_{\text{rel}}$ is the Hamiltonian for the
relative motion of the electrons. 
Kohn's theorem stems from the observation that the electric field is
only coupled directly to the center-of-mass motion and the center-of-mass motion is only coupled to the relative motion through
${\tilde U}({\bf R}_{\text{c.m.}},{\bf r}_{\text{rel},1},\ldots,{\bf r}_{\text{rel},N-1})$.
If ${\tilde U}({\bf R}_{\text{c.m.}},{\bf r}_{\text{rel},1},\ldots,{\bf r}_{\text{rel},N-1})
= {\tilde U}_{\text{c.m.}}({\bf R}_{\text{c.m.}}) + {\tilde U}_{\text{rel}}({\bf r}_{\text{rel},1},\ldots,{\bf r}_{\text{rel},N-1})$, as is the case for
a quadratic confining potential (and for a translationally-invariant system),
the center-of-mass motion
decouples from the relative motion. In such a case,
the response to an electric field is determined entirely by the 
center-of-mass motion.

In a system in its ground state in a quadratic potential, the electric field can only
cause a transition to the first excited state,
so there will be only a single peak in the absorption spectrum. Hence, we must conclude that in our effective theory of the edge, the edge mode velocities and inter-mode interactions are such that the there is only a single charged mode (of a type that we
discuss in the next section). But, if $U$ is not quadratic, the center-of-mass coordinate is coupled to the relative coordinates, and there will be peaks corresponding
to excitations of the relative motion of the electrons. Thus, we will be able
to learn more about the details of the edge structure electromagnetically.

Generically, we do not know the coupling strength between the center-of-mass coordinate and the relative coordinates and the confining potential might have to be tuned in order to see multiple peaks. We expect this coupling to be tunable by changing the shape of the droplet or the steepness of the edge.

\section{Circular droplet}
\label{sec:circle}

\subsection{Laughlin states and $\nu = 1$}\label{sec:laughlin}

\begin{figure}\centering
	\subfloat[Circular droplet without QPC]{
		\centering
		\begin{tikzpicture}
			\draw[fill=gray!20,thick](0,0) circle(1cm);
			\node at (0,0) {$\nu$};
			\draw[->](.84,.84) arc(45:90:1.2cm) node[above right]{$s$};
			\node at (1,0) [right]{$s=0$};
			\node at (1,0) [shape=circle,fill=black,scale=.5] { };
			\draw[<-][thick](-1.5,.5)--(-1.5,-.5) node[below]{$\vec{E}$};
		\end{tikzpicture}
		\label{fig:circle}
	}
	\subfloat[Generic droplet with QPC]{
		\centering
		\begin{tikzpicture}
			\draw[fill=gray!20,rounded corners] (0,1) .. controls(0,0).. (1,0)--(2,0)--(2.1,.9)--(2.2,0)--					(3.2,0)..controls(4.2,0)..(4.2,1)..controls(4.2,2)..(3.2,2)--(2.2,2)--(2.1,1.1)--(2,2)--					(1,2)..controls(0,2)..(0,1);
			\node at (.7,1.3) {$\nu$};
			\node at (4.2,1) [left]{$s=0$};
			\node at (4.2,1) [shape=circle,fill=black,scale=.5] { };
			\node at (2.1,1.1) [right]{$s_a$};
			\node at (2.1,1.25) [shape=circle,fill=black,scale=.5] { };
			\node at (2.1,.9) [below right]{$s_b$};
			\node at (2.1,.75) [shape=circle,fill=black,scale=.5] { };
			\node at (4.3,2.1) {$s$};
			\draw[->](4.4,1.4) arc(0:90:.8);
		\end{tikzpicture}
		\label{fig:QPC}
	}
	\caption{Quantum Hall droplets}\label{fig:droplets}
\end{figure}
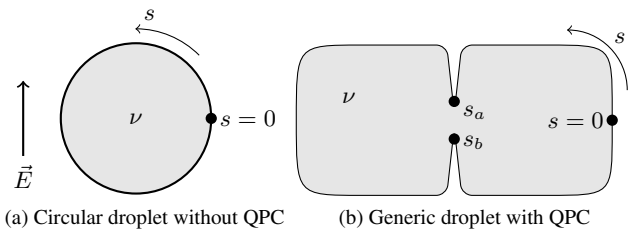

We first consider the simple set-up depicted in Fig~\ref{fig:circle}, a circular quantum Hall droplet in a uniform electric field at filling fraction $\nu = 1/m$, for $m \geq 1$ an odd integer. For these fractions, there is a single edge mode that will couple to the electric field when the frequency of the field matches that of an excitation at the edge. We compute the absorption spectrum using a framework that is generalized in subsequent sections.

The edge modes are described by the chiral Luttinger liquid action\cite{Wen,ChamonFreedWen1,ChamonFreedWen2}
\begin{equation}
\label{eq:L_0}
{\cal S}_0 = \frac{m}{4\pi} \int dt\,ds\,(\partial_t - v\partial_s)\phi(s,t) \partial_s \phi(s,t)
\end{equation}
where $s$ parametrizes the distance along the edge of the droplet, and $v$ is the velocity of the edge mode. The field $\phi$ satisfies the equal-time commutation relation $[\phi(x,t),\partial_s \phi(y,t)] = \frac{2\pi}{m} i\delta(x-y)$. The charge density at a point $s$ along the edge is given by $\rho = \partial_s \phi/(2\pi)$. The electric field of the microwaves
$\vec{E} = E{\cos}(\omega t)\hat{y}$ couples to the charge density of the droplet according to
\begin{equation}
\label{eq:L_E}
\mathcal{L}_E =  E{\cos}(\omega t)y(s)\rho(s,t)
\end{equation}
where $y(s)$ gives the $y$-component of the edge of a droplet; $\rho$ is the
charge density at the edge, given by $\rho = \partial_s \phi/(2\pi)$, where $\phi$ is governed
by the action (\ref{eq:L_0}). For a circular droplet of circumference
$L$, $y(s)=\frac{L}{2\pi}{\rm sin}\left(2\pi s/L\right)$.

This is the minimal edge theory dictated by the bulk qantum Hall state. There can be additional non-chiral
pairs of edge modes, depending on how soft the edge potential is. We will focus here on the case in which there
are only the minimal edge modes dictated by the bulk. The more general case can be analyzed by a straightforward
extension of the present discussion.

The spectrum $R(\omega)$ including both absorption and emission components is found by Fermi's Golden Rule:
\begin{equation}\label{eq:R0}
R(\omega) = \frac{E^2}{2}\int ds_1 ds_2 y(s_1)y(s_2)S^{\rho\rho}(s_1,s_2,\omega)
\end{equation}
where,
\begin{equation}\label{eq:S0}
S^{\rho\rho}(s_1,s_2,\omega) = \int dt {\cos}(\omega t) \langle \rho(s_1,t)\rho(s_2,0)\rangle
\end{equation}
There is a subtlety in computing the $\langle \rho\rho\rangle$ correlation function: because electrons acquire a phase upon circling the droplet, the field $\phi$ is not periodic. However, this phase drops out of all calculations until we include a QPC, so we defer discussion of this phase to Appendix~\ref{sec:interferometrycalc} and here compute the density-density function using the form of $y(s)$ given above and the action (\ref{eq:L_0}). The result is a spectrum with a single pair of peaks at $\omega = \pm 2\pi v/L$:
\begin{equation}\label{eq:Rzerowidth}
R(\omega)  = \nu \frac{E^2 L^2}{32\pi} \delta(\omega \pm 2\pi v/L) 
\end{equation}
Fortunately, these peaks are expected to be in an experimentally accessible regime: using the value $v=10^4$ m/s extrapolated from measurements in Ref~\onlinecite{McClureVelocity}, for a large Hall droplet with $L=50$ $\mu$m the peaks are at frequency $\omega/2\pi = 200$ MHz. The frequency increases inversely with $L$ as the droplet gets smaller in size; for $L=10$ $\mu$m, $\omega/2\pi \approx 1$ GHz. 

\begin{figure}[t]\centering
	\subfloat[Circular droplet]{
		\centering
		\includegraphics[width=.4\textwidth]{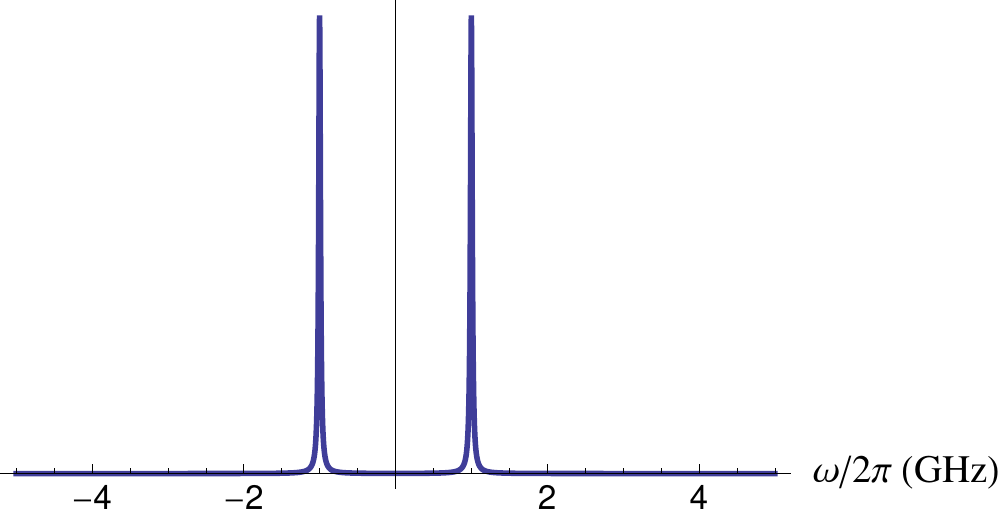}
		\label{fig:SpectrumCircle}
	}
	
	\subfloat[Droplet with QPC]{
		\centering
		\includegraphics[width=.4\textwidth]{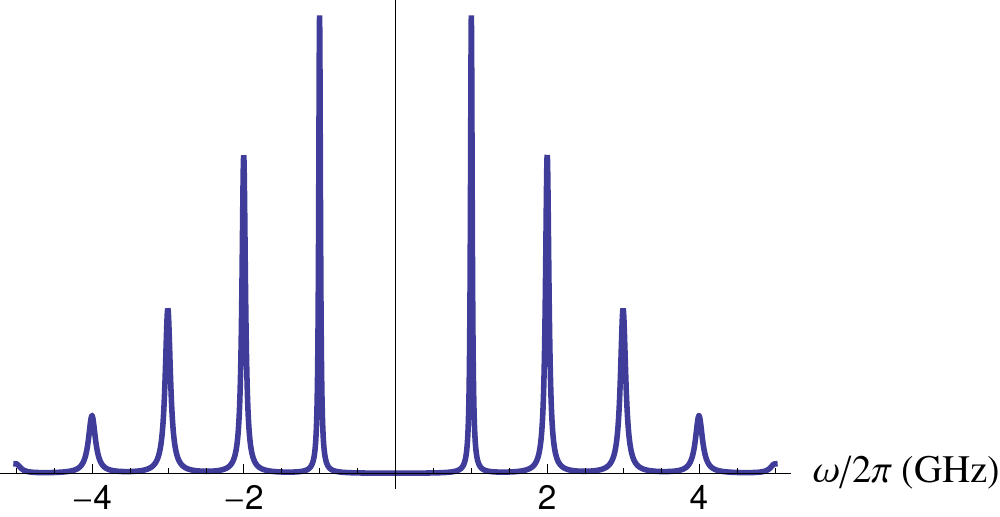}
		\label{fig:SpectrumDoubleCircle}
	}
	\caption{Absorption spectra for a $10\mu$m droplet with $v=10^4$m/s and peak width determined by phonon coupling. The circular droplet (a) permits only one set of peaks, while the non-circular droplet (b) allows a series of peaks.}
\label{fig:Spectrum}
\end{figure}

The delta-function shape of the peaks in the absorption spectrum comes from the isolated poles of the density-density propagator, which correspond to an infinite lifetime for edge excitations. Realistically, the edge excitations will have a finite lifetime due to physics that is neglected in the action of Eq~\ref{eq:L_0}, such as losses in the waveguide, finite longitudinal resistance, and phonon coupling. In a lossless waveguide, phonon coupling will be the dominant contribution to the width and we consider it in detail in Appendix~\ref{sec:peakwidth}. The result is that the spectrum of Eq~\ref{eq:Rzerowidth} is modified to
\begin{equation}\label{eq:Rcircle}
R(\omega) = \nu \frac{E^2 L^2}{32\pi^2} \frac{\eta(\omega)}{(\omega-2\pi v/L)^2+\eta(\omega)^2}
\end{equation}
where $\eta(\omega) = {\rm Im}\left[ \Sigma(k,\omega) \right]$ in Appendix~\ref{sec:peakwidth}. When the piezoelectric contribution dominates that of the deformation potential, as for GaAs, $\eta \propto 1/L$ and the Q-factor of the device is independent of its circumference. For a GaAs device with $v=10^5$m/s, we find $Q \approx 350$. For droplets with circumference $10-50$ $\mu$m, $\eta(\omega) \approx \frac{n}{m}\times 10 {\rm MHz} $ for the $n$th peak at filling fraction $\nu = 1/m$. The absorption spectrum for the circular droplet is shown in Fig~\ref{fig:SpectrumCircle}.

Having only a single peak is specific to the circular droplet. Generically, there are peaks at $\omega = 2\pi n v/L$ for all integers $n$ with width given by $\eta(\omega)$, and the spectrum is generalized for a droplet of arbitrary shape to
\begin{equation}\label{eq:Rgeneral}
R(\omega) = \frac{\nu E^2}{4\pi L}\sum_k \frac{k \eta(\omega)}{(\omega-kv)^2+\eta(\omega)^2}y(k)y(-k)
\end{equation}
This is consistent with the discussion of Kohn's theorem in Sec~\ref{sec:Kohn}. Because a quadratic confining potential can only result in a circular (or elliptical) droplet, these shapes must only have one peak in the absorption spectrum. Other shapes result from non-quadratic terms in the confining potential and are not violating Kohn's theorem by having multiple peaks. For most circle-like shapes, the additional peaks are very small, but for the double-lobed droplet depicted schematically in Fig~\ref{fig:QPC}, several peaks should be visible, as shown in Fig~\ref{fig:SpectrumDoubleCircle}.

At finite temperature there is a prefactor ${\rm coth}(\beta\omega/2)$ to Eqs~\ref{eq:Rcircle} and \ref{eq:Rgeneral}. We do not consider the temperature dependence of $\eta(\omega)$. 

For a droplet of any shape, the absorption spectrum provides a direct measurement of the edge mode velocity. The edge excitation velocity has been deduced in several experiments
at specific filling fractions\cite{AshooriVelocity,KamataVelocity,RezayiVelocity} but there is only one reported measurement that studies its evolution with magnetic field.\cite{McClureVelocity} Measuring the edge velocity from the absorption spectrum would provide a more direct measurement than Ref~\onlinecite{McClureVelocity} and confirm their estimate of when the velocity switches from a `skipping orbit model' to the $\vec{E}\times\vec{B}$ drift velocity.

\subsection{Probing the structure of the edge}\label{sec:many}

When $\nu \neq 1/m$, the edge structure is more complicated. There are expected to be multiple edge modes, which have distinct velocities and mix via density-density interactions and impurity scattering. Here we consider the absorption spectrum of a droplet at a filling fraction with multiple edge modes, first in the clean limit, and then with disorder. 

A perfectly clean edge with $n$ edge modes may be described by the Lagrangian
\begin{equation}
\label{eq:L_many0} \mathcal{L}_{K} = \frac{1}{4\pi} \sum_{ij}\partial_s  \tilde{\phi}_{i}\left( K_{ij} \partial_t + \tilde{v}_{ij}\partial_s \right)\tilde{\phi}_j
\end{equation}
where the $n$-by-$n$ matrix $K$ determines the filling fraction by $\nu = \sum_{ij}K_{ij}^{-1}$ and $v$ is a matrix of non-universal velocities and density-density interactions. The charge density is given by \begin{equation}\label{eq:rhomany} \rho = \sum_i \tilde{\rho}_i \equiv \frac{1}{2\pi}\sum_i \partial_s \tilde{\phi}_i\end{equation}
Following Ref~\onlinecite{KaneFisherGeneral}, we
simultaneously diagonalize $K$ and $v$ by conjugation with a matrix $M$: $\left( M^T \tilde{v}M\right)_{ij} = v_i \delta_{ij} $ and $\left( M^T K M\right)_{ij} = \eta_i \delta_{ij}$ with $\eta_i \in \{ \pm 1 \}$. Then Eq~\ref{eq:L_many0} can be rewritten as a sum of non-interacting chiral edge modes $\phi_i=M^{-1}_{ij}\tilde{\phi}_j$ with respective velocities $v_i\eta_i$, which might be positive or negative:
\begin{equation}
\mathcal{L}_{many} = \frac{1}{4\pi} \sum_{i}\partial_s  \phi_{i}\left( \eta_i \partial_t + v_i \partial_s \right)\phi_i
\end{equation}
Then Eqs~(\ref{eq:L_E})-(\ref{eq:S0}) can be used, and the absorption spectrum
is given by Eq~\ref{eq:R0} with
\begin{multline}
S^{\rho\rho}_{K}(s_1,s_2,\omega) =\\ \int dt {\cos}(\omega t) \sum_{ijk}M_{ij}M^T_{jk}\langle \rho_j(s_1,t)\rho_j(s_2,0)\rangle
\end{multline}
where $\rho_j = \partial_s\phi_j/(2\pi)$. For a circular droplet, we find:
\begin{equation}
\label{eq:Rcirclemany}
R_{K}(\omega) = \frac{E^2L^2}{32\pi}
\sum_{ijm}M_{ij}M^T_{jm} \delta(\omega \pm 2\pi v_j/L)
\end{equation}
We check this result in two simplifying cases: 
\begin{enumerate}
\item When there is only one edge mode, $K=1/\nu$, $M=\sqrt{\nu}$ and Eq~\ref{eq:Rcirclemany} is exactly Eq~\ref{eq:Rzerowidth}.
\item For integer quantum Hall states $\nu = n$ without any inter-mode
density-density interactions, $K=M=\mathbf{I}_n$ and $\tilde{v}$
is diagonal, but generically not proportional to $\mathbf{I}_n$. Then Eq~\ref{eq:Rcirclemany} simplifies to
\begin{equation}R_{\nu =n}(\omega) = \frac{E^2L^2}{32\pi}\sum_{j=1}^n\delta(\omega \pm 2\pi v_j/L) \end{equation}
The spectrum consists of $n$ peaks corresponding to the $n$ different velocities of the edge modes. However, as discussed in Section \ref{sec:Kohn}, if the confining
potential is quadratic, then there will be a single peak, corresponding
to the center-of-mass motion of the entire electron system and
all of the edge mode velocities must be the same.
\end{enumerate}
For generic filling fractions with $n$ edge modes and inter-mode interactions, we expect to see $n$ peaks with a non-universal pre-factor $\sum_{im}M_{ij}M_{mj}$ in front of the $j^\text{th}$ peak. Hence, in the clean limit, the absorption spectrum counts the number of distinct edge modes. If this limit could be observed, it would be direct evidence of the physical reality of the edge mode theory. 

For a droplet of arbitrary shape, Eq~\ref{eq:Rcirclemany} generalizes to
\begin{equation}\label{eq:Rmany} R_{many}(\omega) = \frac{E^2}{4L}\sum_{ijm}M_{ij}M^T_{jm}\sum_k k\hspace{2pt}y(k)y(-k) \delta(\omega - kv_j) \end{equation}
which is the many-mode equivalent of Eq~\ref{eq:Rgeneral}. In this case, we see $n$ series of peaks for which the spacing between peaks in each series is proportional to the velocity of the corresponding edge mode.

The presence of disorder allows equilibration between edge modes, which can dramatically change the edge structure. In Ref~\onlinecite{KaneFisherGeneral}, it is shown that tunneling between edge modes due to scattering off random impurities can drive the system to a random fixed point. For a certain hierarchy of states with $n$ edge modes, the fixed point is stable and has one charged mode and $n-1$ neutral modes. Since the neutral modes do not couple to the electric field, the absorption spectrum in this limit will be exactly that of Sec~\ref{sec:laughlin}: a single peak for a circular droplet or a single series of peaks for a droplet of arbitrary shape. Specifically, at integer fillings $\nu = n$, arbitrarily weak disorder is a relevant term that will drive the edge modes to equilibrate and the absorption spectrum will consist of only one peak corresponding to the equilibrated charged mode. For filling fractions with counter propagating edge modes, such as $\nu = 2/3$, it takes a critical amount of disorder to drive the system into the equilibrated state with only one charged mode. Hence, for weak disorder or for a droplet smaller than the equilibration length, we would expect to see $n$ peaks in the absorption spectrum, as in the clean limit, but when the size of the droplet exceeds the equilibration length, we expect to see only one peak.
Recent experiments detecting neutral upstream modes at expected fractions \cite{ObserveNeutral,LocalThermometry} are presumably in the equilibrated
regime. By considering droplets of multiple sizes (perhaps tuned by gating)
in our setup, both regimes could be observed. Note that, unlike in transport
experiments, we would not have to change the locations of contacts in
order to access different regimes -- there are no contacts in our device.

\section{Interferometry}\label{sec:interferometry}

When probed through microwave absorption, a quantum Hall
droplet with a {\it single} QPC acts an interferometer whose
interference pattern appears as a correction to the height of the
absorption peaks that oscillates with magnetic field.
We will calculate this correction to first-order in the tunneling amplitude and find its dependence on the magnetic field and the number of quasiparticles in the droplet. It is notable that the result is non-zero already at first-order in the tunneling
amplitude, since transport through a Fabry-Perot interferometer would
only see oscillations at second-order in the tunneling amplitude.\cite{ChamonInterferometry,Fradkin98,DasSarma05,Bonderson06a,Stern06,McClureVelocity,McClureInterferometry,ZhangInterferometry,AnInterferometry,LinInterferometry} 

In this section, we consider the cases $\nu = n$ and $\nu = n+1/m$. We model the QPC by adding a tunneling term to the Lagrangian,
\begin{equation}\label{eq:L_{tun}}
\mathcal{L}_{tun} = \lambda e^{i\phi(s_a,t)}e^{-i\phi(s_b,t)} + h.c.
\end{equation}
where in the integer case Eq~\ref{eq:L_{tun}} represents the tunneling of electrons across the QPC while, in the Laughlin case, the term represents the tunneling of charge $e/m$ quasiparticles across the QPC. In the latter case, we could also add a term
to represent the tunneling of electrons across the QPC, but such a term is less relevant.
The position of the QPC is given by $s_a$ and $s_b$,
as shown schematically in Fig~\ref{fig:QPC}.

We want to find $\delta R(\omega)$, the leading order correction to the absorption spectrum in the presence of tunneling. We calculate $\delta R(\omega)$ in Appendix~\ref{sec:interferometrycalc} for a droplet of arbitrary shape. Here we consider a simplified, but realistic, case in which the droplet is symmetrical over the x-axis, so that $y(s) = -y(L-s)$ and $s_b = L-s_a$, yielding
\begin{equation}\label{eq:deltaR}
\delta R(\omega) = 4|\lambda|\!\left[ \frac{E^2}{m^2}{\rm coth}\frac{\beta\omega}{2}   H(\omega) G(\beta)\right]  {\cos}(\varphi)
\end{equation}
where $H$ and $G$ are given by,
\begin{align}\label{eq:HandG}
H(\omega) &= \frac{1}{L^2}\sum_{\substack{ k_{1,2}=2\pi n/L \\ k_1 \neq - k_2}}y(k_1)y(k_2) \frac{{\rm sin}(k_1s_a){\rm sin}(k_2s_a)}{v(k_1+k_2)}\nonumber\\
&\quad\quad\times \left( \frac{\eta(\omega)}{\eta(\omega)^2+ (\omega+k_1v)^2} - \frac{\eta(\omega)}{\eta(\omega)^2+(\omega-k_2v)^2} \right)\end{align}
\begin{align}\label{eq:G}
G(\beta) &= {\rm exp}\left[ \frac{\pi}{mL}\sum_{k>0}\frac{2}{k}({\cos}(2ks_a) -1){\rm coth}\frac{\beta v k}{2} \right] 
\end{align}
$H$ determines the size of the corrections as a function of frequency and
$G(\beta)\coth(\beta\omega/2)$ contains all the temperature dependence of the corrections. Both $H$ and $G$ depend on the placement of the QPC.
The phase $\varphi$ is given by 
\begin{equation}\label{eq:phase}
\varphi = \frac{2\pi}{m}\left( \frac{\Phi_R}{\Phi_0} +n_R -\frac{2s_a}{L}\left( \frac{\Phi}{\Phi_0} + n_{tot} \right)\right) + \alpha\end{equation}
 where $m=1$ for integer states and $m = 1/(\nu - \lfloor \nu \rfloor)$ for Laughlin states, $\Phi$ is the flux penetrating the bulk, $\Phi_R$ is the flux penetrating the right lobe, $n_{tot}$ is the number of quasiparticles in the bulk, $n_R$ is the number of quasiparticles in the right lobe, and $\alpha$ is a phase that is independent of magnetic field. Eq~\ref{eq:phase} shows that for a droplet of fixed shape, the correction $\delta R(\omega)$ varies sinusoidally with magnetic field and its phase is determined by the number of quasiparticles in each lobe.
The basic physical picture is the following. The density-density correlation function
involves the creation and annihilation of a quasiparticle-quasihole pair. Since the
density is integrated over the edge of the droplet, this pair can be created
anywhere. At first-order in the tunneling, the pair can encircle either lobe (which
involves a single tunneling event at the point contact). These different processes will
interfere with each other, and the interference will essentially be controlled by the
{\it difference} between the phases associated with encircling either droplet.
However, the sizes of the lobes matter: it is easier for the pair to encircle
a smaller lobe, so a quasiparticle in a smaller lobe gives a larger contribution to the 
interference phase than a quasiparticle in a larger lobe.
In the next subsections, we will analyze the oscillations and phase shifts; determine the optimal placement of the QPC to see maximum oscillations; and calculate the decay of oscillations at finite temperature.

\subsection{Oscillations and phase shifts}

We now consider the oscillations coming from the phase $\varphi$ in Eq~\ref{eq:phase}. There are two predictions: first, if we fix the number of quasiparticles and vary the magnetic field, we expect oscillations with period proportional to the charge of the quasiparticles:
\begin{equation}
\label{eq:DeltaB}
\Delta B = m\Phi_0 \left( A_R - \frac{2s_a}{L}A \right)^{-1}
\end{equation}
where $A$ and $A_R$ denote the areas of the total droplet and the right lobe, respectively. The factor of two comes directly from Eq~\ref{eq:phase}. We consider this expression in a few limiting cases: first when $s_a \rightarrow 0$, the right side of the droplet disappears so that $A_R \rightarrow 0$. Hence, $\Delta B\rightarrow \infty$ and there are no oscillations; this is what we would expect because the QPC effectively disappears into the right side. Similarly, when $s_a = L/2$, the left side disappears and $A=A_R$; again there are no oscillations and $\Delta B\rightarrow\infty$. The third case is when $s_a = L/4$ and the right and left lobes have equal area $A_R = A_L = A/2$. Then, the left and right lobes enter symmetrically into Eq~\ref{eq:phase}, except for a negative sign. The sign results from the fact that when a particle tunnels, it skips the right lobe but traverses the left lobe one extra time (or vice versa), causing the phases of each lobe enter oppositely. In this case, Eq~\ref{eq:DeltaB} shows that again, oscillations disappear. 

The second prediction is that phase shifts occur for the Laughlin states when the quasiparticle number in either lobe changes, and the phase shift might differ for each lobe. When a quasiparticle is added to the left lobe, the phase shift is $\Delta\varphi = \frac{4\pi}{m}\frac{s_a}{L}$, but when a quasiparticle is added to the right lobe, the phase shift is $\Delta\varphi =  \frac{2\pi}{m}\left(1-\frac{2s_a}{L} \right)$. There will also be a phase shift $\Delta \varphi = \frac{2\pi}{m}$ if a quasiparticle moves from the left to the right lobe. One simplifying case is when the droplet has symmetry about the $y$-axis and $s_a = L/4$: in this case both phase shifts are $\pi/m$ and the oscillations in magnetic field disappear. Without oscillations in magnetic field, it might be easier to observe the statistical phase shift.

This interferometer has the same basic features as the scheme proposed in Ref~\onlinecite{ChamonInterferometry} and executed in Refs~\onlinecite{McClureVelocity,McClureInterferometry,ZhangInterferometry,AnInterferometry,LinInterferometry}, where $\Delta B = mA/\Phi_0$ and $\Delta\varphi = 2\pi/m$ always. However,
our scheme has the additional feature that there is a different phase shift when a quasiparticle is added to the right lobe compared to when one is added to the left lobe,
which makes it possible to see where quasiparticles are added when magnetic flux is
varied. Moreover, it is possible to disentangle the electromagnetic Aharonov-Bohm
effect due to the magnetic flux from the effect of quasiparticle braiding statistics.
For instance, when ${A_R}={A_L}$ (which also means that ${s_a}=L/4$),
changing the magnetic field has no effect whatsoever on the electromagnetic
Aharonov-Bohm phase difference between trajectories encircling the left and right
lobes. However, a change in the magnetic field may result in the creation
of a quasiparticle which will be in either the right lobe or the left lobe
(unless the electrostatics of the device causes us to be in the unlucky
situation in which the quasiparticle sites right at the point contact),
which will lead to a change in the interference phase $\varphi$
of, respectively $\pi/m$ or $-\pi/m$.
The complication is that the phase shifts are non-universal
and depend on the ratio $s_a/L$.

\subsection{Amplitude of oscillations at low temperature}

Next we want to determine where to place the QPC to maximize the amplitude
of oscillations. Both $H$ and $G$ depend on $s_a$; we first consider $H$. If we assume $\eta \ll 2\pi v/L$, then $H$ takes the simplified form when evaluated at the center of a peak,
\begin{equation} \label{eq:Hsum}
H(\omega = 2\pi nv/L) = \frac{2y(k){\rm sin}(ks_a)}{L^2\eta(\omega) v}  \sum_{k'\neq k} \frac{y(k'){\rm sin}(k's_a)}{k'-k}
\end{equation}
where $k = \omega/v$. For simplicity, we consider a droplet whose lobes are perfect circles of radius $R_1$ and $R_2$, and assume that the qualitative features of any droplet with two rounded lobes are captured by this double-circle shape. We then define the ratio $f = s_a/L = R_1/(R_1 + R_2)$, which specifies the position of the QPC. To find the optimal position of $s_a$, we evaluate numerically the dimensionless function $H_n(f) \equiv \left( \eta(\omega) v(2\pi)^5/L^3\right) H(2\pi n v/L)$, shown in Fig~\ref{fig:OscillationMagnitude} for the first few peaks. For the $n=1$ peak, the oscillations are largest for circles of differing radii, but remain sizable throughout the region $.1 < f < .4$.

We now consider $G$ in the low temperature limit:
\begin{align} G(\beta \gg L/v\pi) = \frac{e^{-\gamma/m}}{\left(\frac{2L}{a}{\rm sin}(2\pi f)\right)^{1/m}}
\end{align}
where we have introduced a short-distance cutoff $a$,
and $\gamma$ is the Euler-Mascheroni constant.
In Fig~\ref{fig:GPlot}, we plot the ratio $G(f)/G(f=.25)$ for $\nu = 1, 1/3, 1/5, 1/7$. $G$ is at minimum for the symmetrical droplet and diverges as the droplet reaches maximum asymmetry. Hence, barring the ability to know precisely the shape of the droplet and evaluate Eq~\ref{eq:deltaR} explicitly, an experiment would have to test a variety of asymmetric droplets to find the optimal shape that maximizes the product of $H$ and $G$.

\subsection{Decay of oscillations with temperature}
\begin{figure}[h]\centering
	\subfloat[$H_n(f)$ for various peaks]{
		\centering
		\includegraphics[width=.35\textwidth]{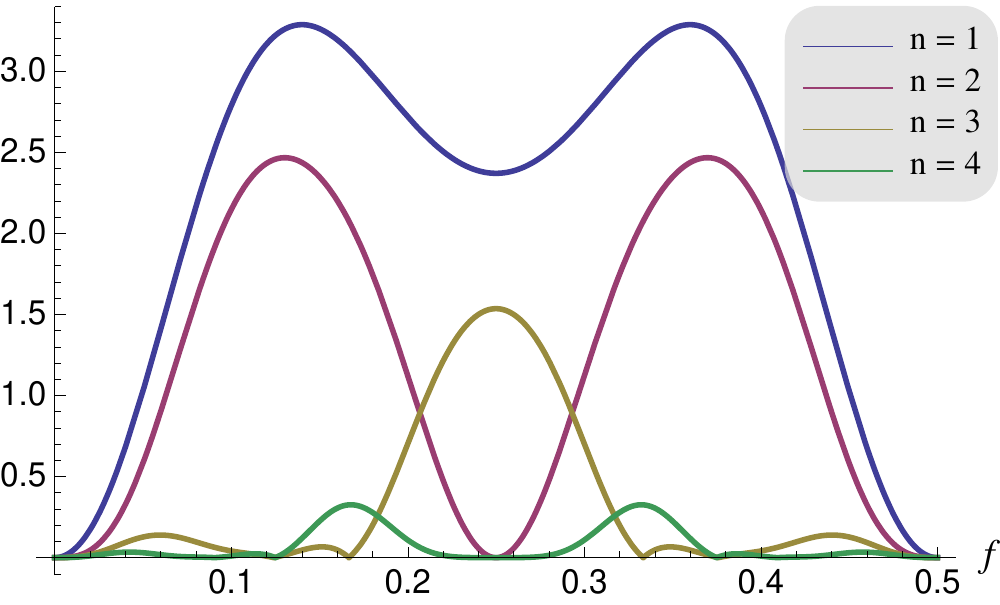}
		\label{fig:OscillationMagnitude}
	}
	
	\subfloat[$G(f)/G(f=.25)$ in the low temperature limit]{
		\centering
		\includegraphics[width=.35\textwidth]{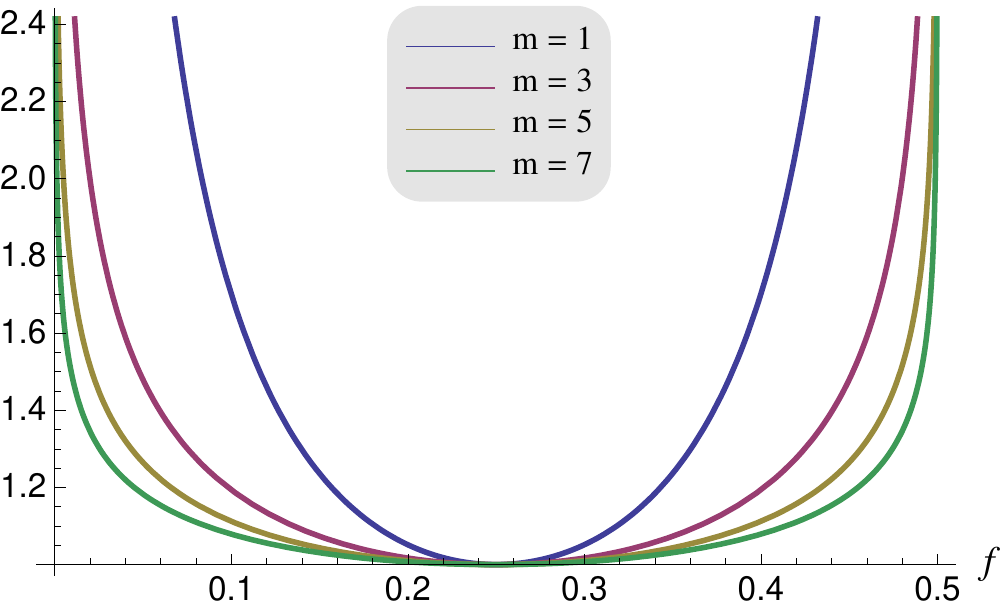}
		\label{fig:GPlot}
	}
	
	\subfloat[Ratio $G(\beta)/G(\beta \rightarrow \infty)$ for $f=.25$ at $\nu = 1/3$]{
		\centering
		\includegraphics[width=.35\textwidth]{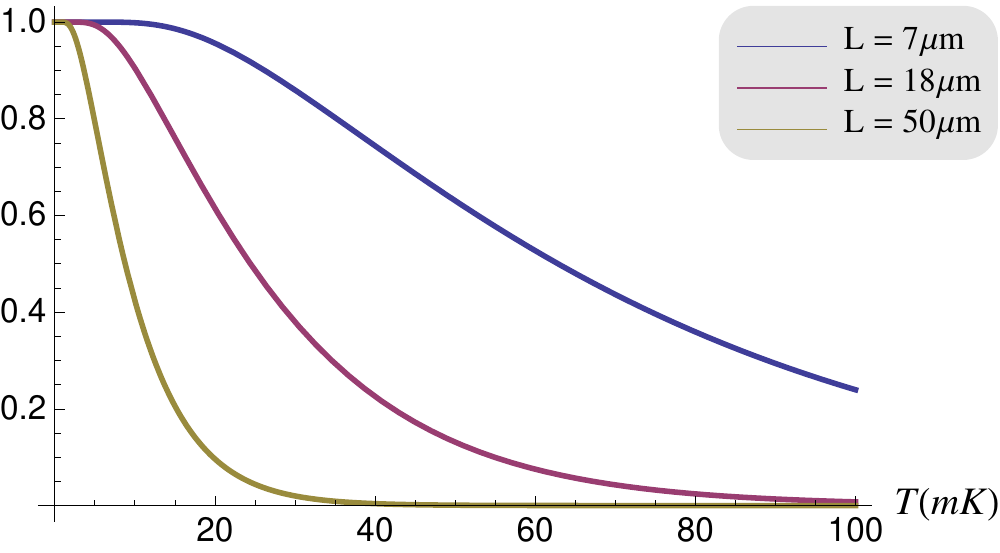}
		\label{fig:GDecay}
	}
	\caption{Dependence of $H$ and $G$ on $f$ and $T$}\label{fig:fdependence}
\end{figure}

In the previous section, we estimated the magnitude of oscillations in the limit $\pi \beta v/L \gg 1$, when the temperature dependence dropped out. This is the correct limit for a very small droplet with $L=2\mu$m, for which the inequality is satisfied for $T \ll 120$mK,
but for intermediate droplets $L=10-50\mu$m, we are not likely to be in this regime. Here we consider the opposite limit when $\beta v\pi/L \ll 1$. In this case, the temperature will define a length scale $L_T$ above which the magnitude of oscillations decays exponentially through the function $G \propto e^{-L/L_T}$. We can roughly estimate the scale of decay by taking only the first term in the sum over $k$ in Eq~\ref{eq:G}, yielding 
\begin{equation} L_T = \frac{\beta \pi v m}{ 1-\cos(4\pi f) } \geq \beta \pi v m/2 \equiv L_T^{min}
\end{equation}
The symmetric droplet ($f=\frac{1}{4}$) achieves the minimum limit $L_T = L_T^{min}$. The length scale diverges for the maximally asymmetric droplets with $f=0$ or $f=.5$. Taking $v=10^4$m/s and $\nu = 1/3$, $L_T^{min} = 7\mu$m at 50mK and $L_T^{min} = 18\mu$m at 20mK. This temperature dependence is shown in Fig~\ref{fig:GDecay} at $\nu = 1/3$ for the symmetric droplet with $f=.25$. For asymmetric shapes, the droplet could be larger.

\section{Non-Abelian interferometry at $\nu = 5/2$}

Recent experiments\cite{Radu,Upstream,Willett09,Willett10,Willett13} suggest that the wavefunction of a quantum Hall state at filling fraction $\nu = 5/2$ is in the same universality class as the Moore-Read Pfaffian state\cite{Moore91,Greiter92,Nayak96c} or the
anti-Pfaffian state\cite{Lee07,Levin07}. Both states have non-Abelian quasiparticles
of Ising type\cite{Bonderson11a}.
Here we will determine the absorption spectrum for the Pfaffian and anti-Pfaffian states,
first without and then with a QPC. We show that the latter system acts as a
non-Abelian interferometer. 

\subsection{Droplet with no QPC}
The edge theories of the Pfaffian and anti-Pfaffian states consist of a charged chiral boson $\phi$ and a chiral Majorana fermion, $\psi$. The fermion is the charged part of the critical 2D Ising model, and is accompanied by a twist field $\sigma$. To be precise, the anti-Pfaffian state has three chiral Majorana fermions, but since they are not coupled in our model, we will only consider one. In the Pfaffian state, the charged and neutral modes have the same chirality, while in the anti-Pfaffian state, they have opposite chirality, but this difference will not appear in the absorption spectrum. The bosonic edge modes are described by the Lagrangian of Eq~\ref{eq:L_0} with $m=2$, which couple to the electric field through Eq~\ref{eq:L_E}. The neutral fermions do not couple to the electric field and, consequently, do not affect the spectrum of a droplet without a QPC. Hence, the analysis of a droplet without a QPC is identical to that of the integer and Laughlin states. The absorption spectrum for an
arbitrary shape is the same as in Sec~\ref{sec:laughlin}, where it is given in Eq.~\ref{eq:Rgeneral}. There will be peaks at all frequencies that are multiples of $2\pi v/L$ and the peak placement is a direct measurement
of the bosonic edge velocity. At non-zero temperature, there is a pre-factor ${\coth}(\beta \omega/2)$.

Another leading candidate to describe filling fraction $\nu = 5/2$ is the
Abelian (3,3,1) state\cite{Halperin1}. In the limit of a clean edge, this state has a different interferometric signature: following Sec~\ref{sec:many}, we would expect to see two series of peaks in the absorption spectrum corresponding to the two edge modes, in contrast to the single series for the Pfaffian and anti-Pfaffian. However, the states will be indistinguishable if there is disorder on the edge that drives the edge modes to equilibrate. Henceforth, we
will focus on the Pfaffian and anti-Pfaffian states.

\subsection{Non-Abelian interferometry with a QPC}
We see new effects when a QPC is introduced. There are several types of quasiparticles, all of which could tunnel across the QPC, but the most relevant tunneling term is that for charge $e/4$ quasiparticles, given by
\begin{equation}\label{eq:L_tun52}
\mathcal{L}_{tun}^{5/2} = \lambda \Phi_{1/4}(s_a,t)\Phi^\dagger_{1/4}(s_b,t) + h.c.
\end{equation}
where $\Phi_{1/4} = \sigma e^{i\phi/2}$ is the annihilation operator for a charge $e/4$ quasiparticle. In Appendix~\ref{sec:interferometrycalc52}, we detail the method to compute the correction to the absorption spectrum to order $\lambda$. The result for a droplet symmetrical over the x-axis is
\begin{equation}\label{eq:deltaRnonAbelian}
\delta R(\omega) = 4|\lambda|\!\left[ \frac{E^2}{16}{\coth}\frac{\beta \omega}{2}  H(\omega)G(\beta)\right] \left|\mathcal{G}(s_a)\right| \,{\cos}(\varphi_{5/2})
\end{equation}
where $H$ and $G$ are the same as for the integer and Laughlin fractions and are given by Eq~\ref{eq:HandG}-\ref{eq:G} with $m=2$. The phase $\varphi_{5/2}$ is given by
\begin{equation}
\varphi_{5/2} = \frac{\pi}{2} \left( \frac{\Phi_R}{\Phi_0} + \frac{n_R}{2} -\frac{2s_a}{L}\left(\frac{\Phi}{\Phi_0} + \frac{n_{tot}}{2}\right)\right) + \frac{\pi}{16}+ \alpha 
\end{equation}
As in the Abelian case, the interference phase is essentially controlled
by the difference in the phase associated with encircling the right droplet
and the left droplet. The glaring difference between the Abelian and non-Abelian case is the presence of the function 
\begin{equation}
\mathcal{G}(s_a) \equiv \langle \sigma(s_a) \sigma(L-s_a) \rangle
\end{equation}
in Eq~\ref{eq:deltaRnonAbelian}, which is the topological contribution to the phase and will be the focus of the rest of the discussion in this section. $\mathcal{G}$ depends on the total topological charges in the right and left lobes, which we denote by $F_{R/L} \in \lbrace I, \psi,\sigma\rbrace$. When there are no quasiparticles in the bulk, 
\begin{equation}
\mathcal{G}(s_a)_\text{no qp} = \mathcal{G}_0(s_a) \equiv e^{i\pi/16}\left(\frac{L}{\pi} {\rm sin}(2\pi s_a/L)\right)^{-1/8}
\end{equation}
For other topological charges, the result is proportional to $\mathcal{G}_0(s_a)$,
as shown in Table~\ref{tab:sigmasigma}.

Table~\ref{tab:sigmasigma} shows that having quasiparticles in the bulk can reduce the amplitude of oscillations. In particular, when there is an odd number of $\sigma$ quasiparticles in each lobe, oscillations disappear completely. The disappearance is a direct consequence of non-Abelian statistics: if the two bulk quasiparticles are fused to $I$ (or $\psi$), the tunneling quasiparticles will flip their fused state from $I$ to $\psi$ (or vice versa), causing the first order term to disappear. This term will re-appear to second-order from virtual tunneling processes (and might be visible as smaller oscillations). Hence, if the number of quasiparticles is varied, either explicitly or by changing the area of the droplet or the magnetic flux penetrating it, we would expect to see oscillations that disappear when there are an odd number of quasiparticles in each lobe.
This effect was predicted for an interferometer with two QPCs in Ref~\onlinecite{Bonderson06a,Stern06}, has been analyzed in great detail in Ref~\onlinecite{Bonderson2,BondersonLong,InterferometricSignatures,OddEvenCrossover,PfaffvsAntiPfaff} and has been seen experimentally in Ref~\onlinecite{Willett09,Willett10,Willett13}. The experiment we propose would be complementary to existing experiments and has
the advantage that oscillations are first-order in the tunneling amplitude.

\begin{table}
\begin{tabular}{c|c|c}
$F_L$ & $F_R$ & $\mathcal{G}(s_a)/\mathcal{G}_0(s_a)$ \\ \hline 
$I$ & $\psi$ & $-{\cos}(2\pi s_a/L)$ \\
$I$ & $\sigma$ & $\left( {\rm sin}(\pi s_a/L) \right)^{1/2}$\\
$\psi$ & $\psi$ & 1\\
$\psi$ & $\sigma$ & $\left( {\rm sin}(\pi s_a/L) \right)^{1/2}$ \\
$\sigma$ & $\sigma$ & 0
\end{tabular}\caption{Topological pre-factor to peak height}\label{tab:sigmasigma}
\end{table}

Finally, we note that there is a simple statistical mechanical interpretation
of this table. The correlation function $\mathcal{G}(s_a)$ is the expectation
value of the spin $\langle\sigma\rangle$ in the critical
Ising model (either the critical classical $2D$ Ising model or the critical transverse
field quantum Ising model) on a strip of length $L/2$ with
specified boundary conditions. If there are no quasiparticles in the bulk,
the Ising model has fixed $+$ boundary conditions at both ends,
and $\langle\sigma\rangle$ is given by the power-law decay from the ends
characteristic of the critical point. If the total charge (of both lobes combined)
is $\psi$, then the two ends have opposite
fixed boundary conditions, $+$ at one end
and $-$ at the other. Then $\langle\sigma\rangle$ vanishes at the midpoint
of the strip and is either positive or negative to the left or right of the midpoint
(i.e. if the left or right lobe is larger, ${s_a}<L/4$ or ${s_a}>L/4$). If the total
topological charge is $\sigma$, then one end of the Ising model has fixed
boundary conditions and the other free boundary conditions,
and $\langle\sigma\rangle$ is given by the power-law decay from one end
characteristic of the critical point. Finally, if there is topological charge
$\sigma$ in each lobe, then the Ising model has free boundary conditions
at both ends and $\langle\sigma\rangle$ is simply zero.


\section{Discussion}

In this paper we propose a new method to probe the edge of a quantum Hall droplet by measuring its microwave absorption spectrum. For a simple, circular droplet, this measurement would reveal the number of charged modes and their velocities. For edges with counter propagating edge modes, this information would resolve open questions about how current is carried at the edge. When a QPC is introduced, the droplet can serve as an interferometer. Its capabilities are similar to existing proposals and experiments, but has the advantage that the amplitude of oscillations is first-order in the tunneling amplitude.
There are also subtle differences from transport through a two point contact interferometer,
such as a dependence on the side of the QPC to which a quasiparticle
has been added, which leads to a non-universal phase shift. 

At $\nu = 5/2$, such a measurement could determine if the state of the system is non-Abelian: if it is, then oscillations in the absorption
spectrum appear when there are an even number of $\sigma$ quasiparticles
in each lobe but not when there is an odd number.
This experiment would be complementary to existing interferometry
experiments\cite{Willett09} at $\nu = 5/2$ and,
as in the Abelian case, has the advantage of having
oscillations at first-order in the tunneling amplitude.

Thus far, quasiparticle properties of fractional quantum Hall states have
been deduced from resistance oscillations in mesoscopic devices.
Here, we propose a new approach in which this information is gathered
from the absorption spectrum. It could confirm existing experimental results and, in doing so, resolve questions on the fundamental tenets of the theory of the fractional quantum Hall effect.

Moreover, by coupling a quantum Hall device to microwaves,
we open the possibility of using photons as a quantum bus
to transfer information from a $\nu=5/2$ qubit \cite{DasSarma05,Nayak08}
to superconducting or even semiconductor quantum dot qubits.

\subsection{Acknowledgements}
We would like to thank Steve Kivelson, Roman Lutchyn, Michael Mulligan, and
Javad Shabani for helpful discussions and James Colless, John Hornibrook, Alice Mahoney, and Xanthe Croot for experimental details. We gratefully acknowledge the support of the DARPA QuEST
program, and the Australian Research Council Centre of Excellence Scheme (Grant No. EQuS CE110001013). J.C. acknowledges the support of the National Science Foundation Graduate Research Fellowship under Grant No. DGE1144085.

\appendix

\section{Magnetoplasmon lifetime from phonon coupling}
\label{sec:peakwidth}

Here we calculate the decay rate of the magnetoplasmon edge modes at Laughlin fractions $\nu = 1/m$ by considering their coupling to phonons, following Ref. \onlinecite{Zulicke97}
\begin{equation}\label{eq:mp-ph}
\mathcal{S}_{mp-ph} = \int dt d^3 \vec{x}d^3 \vec{x}' \rho_{3D}(\vec{x},t) \mathcal{V}_{ij}(\vec{x}-\vec{x}')\partial_i u_j(\vec{x}',t) \end{equation}
where the field $\vec{u}$ is the ion displacement field, $\rho_{3D}$ is the 3D charge density, and $\mathcal{V}$ is symmetric in its indices and
contains the deformation potential, which is the effect of
deformations of the lattice on the local electron density, and the piezoelectric
effect, which is the long-ranged electric field caused by lattice distortions,
according to (as discussed, for instance, in Ref~\onlinecite{Mahan})
\begin{equation} \mathcal{V}_{ij}(\vec{x}-\vec{x}') = D\delta(\vec{x}-\vec{x}')\delta_{ij} +
e h_{14}V_{ij}(\vec{x}-\vec{x}') \end{equation}
where $D$ is the deformation potential constant, $h_{14}$
is the piezoelectric coupling constant, and the electric potential generated
by a lattice strain satisfies
\begin{equation}\sum_i q_i V_{ij}(\vec{q}) = \sum_\lambda iM_\lambda(\hat{q})\left( \epsilon_{\vec{q}}^\lambda\right)_j \end{equation}
where $\epsilon_{\vec{q}}^\lambda$ is the polarization vector for a phonon with polarization $\lambda$ and wave vector $\vec{q}$, $h_{14}$ is a piezoelectric coupling constant and $M_\lambda$ is an anisotropy factor.

We consider here a circular geometry,
\begin{equation} \rho_{3D} (\vec{x}) = \delta(z)\delta(R-|r|)\rho(s) \end{equation}
for which the interaction action can be rewritten as
\begin{equation}S_{mp-ph} = \int \frac{d\omega}{2\pi}\frac{1}{L}\sum_k \int \frac{d^3\vec{q}}{(2\pi)^3} \phi(k,\omega) u_j(\vec{q},-\omega)C_j(k,\vec{q}) \end{equation}
where 
\begin{equation}C_j(k,\vec{q}) = -q_i k\frac{L}{2\pi} \mathcal{V}_{ij}(\vec{q})e^{in_k\theta_{\vec{q}}}(i)^{n_k} J_{n_k}(\frac{L}{2\pi} q_\parallel) \end{equation}
and $\theta_{\vec{q}}$ is the azimuthal angle of $\vec{q}$; $q_{\parallel}$ is the magnitude of the component of $\vec{q}$ that lies in the plane of the droplet; $J_n$ is the $n$th Bessel function of the first kind; and  $n_k\equiv kL/2\pi$.

The decay rate of the edge modes is given by the self-energy
\begin{align} \Sigma(k,\omega) = -\frac{2\pi i}{Lmk} \int\frac{d^3\vec{q}}{(2\pi)^3} &\langle u_i(\vec{q},\omega)u_j(-\vec{q},-\omega) \rangle \nonumber\\ & \quad C_i(k,\vec{q})C_j(-k,-\vec{q}) \end{align}
The imaginary part of the self-energy gives us the decay rate of the edge mode. Using the phonon propagator,
\begin{equation} \langle u_i(\vec{q},\omega)u_j(-\vec{q},-\omega) \rangle = \frac{1}{\rho}\frac{i(\epsilon_{\vec{q}}^\lambda)_i(\epsilon_{-\vec{q}}^\lambda)_j}{\omega^2 -(v_\lambda q)^2+i\delta}\end{equation}
where $\rho$ is the mass density of the device, we find the imaginary part of the self-energy,
\begin{align}
{\rm Im} & \left[ \Sigma(k,\omega) \right] = \frac{k L}{2\pi m\rho}\int \frac{d^3\vec{q}}{(2\pi)^3}\left( J_{n_k} \left(q_\parallel L/2\pi \right) \right)^2 \nonumber\\
& \left( D^2 q^2 + e^2h_{14}^2\sum_\lambda \left(M_\lambda(\hat{q})\right)^2\right) \pi \delta(\omega^2 -v_\lambda^2 q^2)
\end{align}
We immediately find the contribution proportional to $D^2$:
\begin{equation} {\rm Im}\left[ \Sigma(k,\omega) \right]^D = D^2\frac{k\omega^2}{8\pi m\rho v_l^4} \end{equation}
To find the piezoelectric contribution requires the anisotropy factors for a 2DEG oriented on the (001) plane of GaAs:\cite{Price81}
\begin{align}
\left( M_l(q_\parallel,q_\perp) \right)^2 &= \frac{9q_\perp^2 q_\parallel^4}{2(q_\perp^2 + q_\parallel^2)^3} \nonumber\\
\left( M_t(q_\parallel,q_\perp) \right)^2 &= \frac{8q_\perp^4 q_\parallel^2 + q_\parallel^6}{4(q_\perp^2+q_\parallel^2)^3}
\end{align}
where $q_\perp$ is the component of $\vec{q}$ perpendicular to the plane of the 2DEG, from which we find,
\begin{equation} {\rm Im}\left[ \Sigma(k,\omega) \right]^{pz} = e^2h_{14}^2 \frac{k}{8\pi m\rho} \left( \frac{13}{32}\frac{1}{v_t^2} + \frac{9}{32}\frac{1}{v_l^2} \right) \end{equation}
We now specialize to the parameter values for a GaAs quantum well\cite{Lyo88,Kawamura92,Zulicke97}: $D=12$eV, $h_{14} = 1.2\times 10^7$V/cm, $v_l = 5.14\times 10^3$m/s, $v_t = 3.04\times 10^3$m/s and $\rho = 5.3$g/cm$^3$. For droplets with $L=10-50\mu$m, these numbers yield
\begin{align}
{\rm Im}\left[ \Sigma (k,\omega) \right]^D &=(\nu n_k^3)\times \left( .01 - 1.5 {\rm kHz}\right) \nonumber\\
{\rm Im} \left[ \Sigma(k,\omega)\right]^{pz} &= (\nu n_k) \times \left(2.8 - 14 {\rm MHz} \right)
\end{align}
where $n_k = Lk/2\pi$ is the mode number.

\section{Correction to absorption peak heights in the presence of a QPC for integer and Laughlin states}
\label{sec:interferometrycalc}

To find the first order correction $\delta R(\omega)$  in the presence of $\mathcal{L}_{tun}$  it is helpful to define the retarded Green's function $\chi(s_1,s_2,t) \equiv -i\Theta(t) \langle \left[ \phi(s_1,t),\phi(s_2,0)\right] \rangle$ and its Fourier transform $\chi(k_1,k_2,\omega) \equiv \int_0^L ds_1ds_2 e^{-ik_1s_1-ik_2s_2}\int dt e^{i\omega t} \chi(s_1,s_2,t)$, in terms of which we can write the absorption spectrum using the fluctuation-dissipation theorem
\begin{align}\label{eq:Rfluc-diss}
R(\omega) &= \frac{E^2}{16\pi^2L^2}{\coth}\frac{\beta\omega}{2}\sum_{k_1,k_2}(-k_1k_2) y(k_1)y(k_2)\nonumber\\
&\times   \Big( i\chi(-k_1,-k_2,\omega) -i\chi(-k_2,-k_1,-\omega) \Big)
\end{align}
To find the order $\lambda$ correction to $R(\omega)$ we need to find the order $\lambda$ correction to $\chi$. We do perturbation theory in imaginary time:
\begin{align}\label{eq:ImTimeChi}
\delta\chi&(s_1,s_2,\omega_n) = \int_0^\beta d\tau d\tau' e^{i\omega_n\tau} \Big(  \langle \phi(s_1,\tau)\phi(s_2,0) \mathcal{L}_{tun}(\tau') \rangle_0 \nonumber\\
&\quad\quad -\langle\phi(s_1,\tau)\phi(s_2,0)\rangle_0 \langle \mathcal{L}_{tun}(\tau') \rangle_0 \Big) \nonumber\\
&= \int_0^\beta d\tau d\tau' e^{i\omega_n\tau} \Big(  \langle \phi(s_1,\tau)(\phi(s_a,\tau')-\phi(s_b,\tau')) \rangle_0 \nonumber\\
&\quad\quad \times  \langle \phi(s_2,0)(\phi(s_a,\tau')-\phi(s_b,\tau') )\rangle_0 \langle \mathcal{L}_{tun} (\tau') \rangle_0 \Big) \nonumber\\
&= \left( \chi_0(s_1,s_a,\omega_n)-\chi_0(s_1,s_b,\omega_n) \right) \nonumber\\
&\quad\quad\times \left( \chi_0(s_2,s_a,-\omega_n)-\chi_0(s_2,s_b,-\omega_n) \right) \langle \mathcal{L}_{tun}(0) \rangle_0
\end{align}
where all the expectation values are imaginary time ordered and the subscript 0 indicates correlation functions calculated at $\lambda = 0$. We have omitted contributions to a zero frequency peak. The middle equality comes from the identity for quadratic fields $\langle \hat{O}_1\hat{O}_2e^{i\alpha \hat{O}_3 }\rangle = \left( \langle \hat{O}_1\hat{O}_2 \rangle - \alpha^2 \langle \hat{O}_1\hat{O}_3\rangle \langle\hat{O}_2\hat{O}_3\rangle\right)\langle e^{i\alpha \hat{O}_3 }\rangle$ and time translational invariance allows us to change the argument of $\mathcal{L}_{tun}$ in the last line. Using the Lagrangian~(\ref{eq:L_0}), we find,
 \begin{equation} \chi_0(s_i,s_j,\omega_n) =  -\frac{2\pi}{mL} \sum_{k_j} \frac{1}{k_j} \frac{e^{ik(s_i-s_j)}}{i\omega_n -kv} \end{equation}
 From which we can simplify Eq~\ref{eq:ImTimeChi}
 \begin{align}\delta\chi(s_1,s_2,\omega_n) &= \frac{4\pi^2}{m^2L^2} \prod_{j=1,2}\left( \sum_{k_j} \frac{e^{ik_js_j} \left( e^{-ik_j s_a} - e^{-ik_js_b}\right) }{k_j((-1)^{j+1}(i\omega_n) -k_jv)}\right) \nonumber\\
 &\quad\quad\times \langle \mathcal{L}_{tun}(0)\rangle_0
 \end{align}
 and find the order $\lambda$ correction to Eq~\ref{eq:Rfluc-diss} by taking $i\omega_n \rightarrow \omega+ i\eta$,
\begin{align}\label{eq:deltaR-result}
\delta &R(\omega) = -  \frac{E^2}{4m^2}{\coth}\frac{\beta\omega}{2} \langle \mathcal{L}_{tun}(0)\rangle_0   \frac{1}{L^2}\sum_{k_1,k_2}y(k_1)y(k_2)\nonumber\\
& \times (e^{ik_1s_a}-e^{ik_1s_b})(e^{ik_2s_a}-e^{ik_2s_b}) \nonumber\\
& \times \frac{2}{v(k_1+k_2)}\left( \frac{\eta(\omega)}{\eta(\omega)^2+ (\omega+k_1v)^2} - \frac{\eta(\omega)}{\eta(\omega)^2+(\omega-k_2v)^2} \right)
\end{align}
There is a subtlety in calculating the expectation value of the tunneling Lagrangian $\langle \mathcal{L}_{tun}(0)\rangle_0$. As mentioned in Sec~\ref{sec:laughlin}, electrons acquire a phase $\theta$ upon circling the droplet, so the field $\phi$ is not perfectly periodic. Instead, $\Psi_{el}(s=0) = \Psi_{el}(s=L)e^{i\theta}$, where $\Psi_{el}=e^{im\phi}$ is the electron annihilation operator, from which it follows that $\phi(0) = \phi(L) +\theta/m$. Hence, the mode expansion of $\phi$ includes a zero-mode proportional to $\theta$:
\begin{multline}
\label{eq:quantizephi}
\phi(s,t) =  - \frac{\theta}{m} \frac{s}{L}\\
+ \sqrt{\frac{2\pi}{mL}}\sum_{k=\frac{2\pi n}{L}>0} \frac{1}{\sqrt{k}}\left( e^{ik(s-vt)}\phi_k + e^{-ik(s-vt)}\phi_k^\dagger \right)
\end{multline}
We can regard $\theta$ as a classical variable that commutes with the $\phi_k$, which themselves satisfy $\left[ \phi_k,\phi_{k'}^\dagger \right] = \delta_{k,k'}$. The value of $\theta$ is fixed by macroscopic parameters: 
\begin{equation} \theta = 2\pi (\Phi/\Phi_0 + n_{tot})\end{equation}
where $\Phi$ is the flux penetrating the droplet, $\Phi_0=h/e$ is the flux quantum and $n_{tot}$ is the number of charge $e/m$ quasiparticles in the bulk in the Laughlin case (in the integer case, $n_{tot} = 0$.) Consistency with the definition of $\theta$ requires ${\rm arg}(\lambda) = -\theta_L/m+\alpha$, where $\alpha$ is independent of magnetic field and quasiparticle number and $\theta_L = 2\pi(\Phi_L/\Phi_0 + n_L)$ is the Aharonov Bohm phase an electron would acquire from circling only the left lobe of the droplet when there is flux $\Phi_L$ piercing the left lobe and $n_L$ quasiparticles inside. Using the mode expansion (\ref{eq:quantizephi}), we can now correctly evaluate:
 \begin{align}\label{eq:expL_tun}
\langle &  \mathcal{L}_{tun}(0)\rangle_0 = |\lambda| 2{\cos} \left( {\rm arg}(\lambda)-\frac{(s_a-s_b)\theta}{Lm}\right)\nonumber\\
&\quad \times {\rm exp}\left[ \frac{\pi}{mL} \sum_{k>0}\frac{2}{k}({\cos}(k(s_a-s_b)) -1){\rm coth}\frac{\beta v k}{2} \right]\nonumber\\
\end{align}

\section{Correction to absorption peak heights in the presence of a QPC at $\nu = 5/2$}\label{sec:interferometrycalc52}
To find the first order correction $\delta R(\omega)$ to the absorption spectrum at $\nu = 5/2$  in the presence of $\mathcal{L}^{52}_{tun}$ we follow the calculation in Appendix~\ref{sec:interferometrycalc} and reach Eq~\ref{eq:deltaR-result} with $m =2$ and 
\begin{align}
\langle \mathcal{L}_{tun}&(0)\rangle_0  \rightarrow \langle \mathcal{L}_{tun}^{5/2}(0) \rangle_0\nonumber\\
&= \lambda \langle \sigma(s_a,0)\sigma(s_b,0) \rangle \langle e^{i\phi(s_a,0)/2}e^{-i\phi(s_b,0)/2} \rangle + h.c.
\end{align}
We can find
\begin{align} 
\langle & e^{i\phi(s_a,0)/2} e^{-i\phi(s_b,0)/2} \rangle =\nonumber\\
& {\rm exp}\Bigg[ \frac{\pi}{L} \sum_{k>0} \frac{1}{k}\left( {\cos}(k(s_a-s_b)) -1 \right){\rm coth}\frac{\beta vk}{2} -i\frac{\theta_{5/2}}{4}\frac{s_a-s_b}{L} \Bigg]
\end{align}
where we have used the mode expansion 
\begin{equation}
\label{eq:quantizephi52}
\phi(s,t) = \sqrt{\frac{\pi}{L}}\sum_{k>0} \frac{1}{\sqrt{k}}\left( e^{ik(s-vt)}\phi_k + e^{-ik(s-vt)}\phi_k^\dagger \right) - \frac{\theta_{5/2}}{2} \frac{s}{L}
\end{equation}
where $\theta_{5/2} = 2\pi\left( \Phi/\Phi_0 + n_{tot}/2 \right)$ and $n_{tot}$ is the number of $e/4$ quasiparticles in the bulk, to account for the non-periodicity of $\phi$, as described in Appendix~\ref{sec:interferometrycalc}. Consistency with the choice of $\theta_{5/2}$ requires ${\rm arg}(\lambda) = -\theta_L/4+\alpha$, where $\alpha$ is the non-Aharonov Bohm contribution to the phase and $\theta_L=2\pi(\Phi_L/\Phi_0 + n_L)$ is the Aharonov-Bohm phase an electron would acquire from circling the left lobe of the droplet. 
 
We now seek  the correlation function $\langle \sigma(s_a)\sigma(s_b) \rangle $. When there are no bulk quasiparticles, the correlation function is given by\cite{CFT}
 \begin{equation} \langle \sigma(s_a)\sigma(s_b)\rangle_\text{no qp} =  \frac{e^{-i\pi/16}}{\left( \frac{L}{\pi}{\rm sin}\left( \frac{\pi}{L}(s_a-s_b) \right) \right)^{1/8}}\end{equation}
 A quasiparticle in the bulk contributes a branch cut that crosses the perimeter of the droplet at some point $s_j$ for all times. We can think of this branch cut as coming from the creation of a quasiparticle at $s_j$ at a time in the far past and its subsequent annihilation in the far future. Hence, the two point function with one quasiparticle in the bulk is computed by
 \begin{align} \langle & \sigma(s_a,t) \sigma(s_b(t) \rangle_\text{bulk qp} =\nonumber\\
& \lim_{T\rightarrow\infty} \frac{\langle(\sigma(s_j,T)\sigma(s_a,t)\sigma(s_b,t)\sigma(s_j,-T) \rangle_\text{no qp}}{\langle \sigma(s_j,T)\sigma(s_j,-T) \rangle_\text{no qp} } \end{align}
The numerator and denominator can be calculated using bosonization\cite{CFT}, specifically, by the method of \onlinecite{SigmaCorrelators}. Additional quasiparticles can be included by adding more pairs to the numerator and denominator. When we do this, we always assume that the pair of $\sigma$ quasiparticles at the QPC are fused to the identity, i.e. there is an energy cost for creating a fermion on the edge. We also assume that the fermion parity of the entire system, consisting of the droplet and the point at infinity, is even. We cite the results for specific cases in the main text.

\bibliography{QHBibliography}

\end{document}